\documentclass[pra,aps,showpacs,floatfix,preprint]{revtex4-1}
\usepackage{color}
\usepackage[utf8]{inputenc}

\RequirePackage{amsfonts}

\usepackage{colordvi}

\usepackage{graphicx,amssymb,epsfig,amsmath}
\usepackage{epstopdf}

\newcommand{\be}{\begin{equation}}
\newcommand{\ee}{\end{equation}}
\newcommand{\bea}{\begin{eqnarray}}
\newcommand{\eea}{\end{eqnarray}}

\renewcommand{\a}[1]{\hat{a}_{#1}}
\newcommand{\ad}[1]{\hat{a}_{#1}^\dagger}
\newcommand{\ak}{\hat{a}_k}
\newcommand{\adk}{\hat{a}^\dagger_k}
\newcommand{\aq}{\hat{a}_q}

\newcommand{\ads}{\hat{a}^\dagger_s}
\newcommand{\al}{\hat{a}_l}

\renewcommand{\vr}{\vec{r}}
\newcommand{\Cn}{C_{\vec{n}}(t)}

\begin{document}
\title{The multiconfigurational time-dependent Hartree for fermions: Implementation, exactness and few-fermion tunneling to open space.}

\author{Elke Fasshauer}
\email{elke.fasshauer@uit.no}
\affiliation{Centre for Theoretical and Computational Chemistry, Department of Chemistry, University of Troms\o{} -- The Arctic University of Norway, N-9037 Troms\o{}, Norway}

\author{Axel U. J. Lode}
\email{axel.lode@unibas.ch} 
\affiliation{Department of Physics, University of Basel, Klingelbergstrasse 82, CH-4056 Basel, Switzerland}

\date{\today}

\begin{abstract}
We report on an implementation of the multiconfigurational time-dependent Hartree method (MCTDH) for spin-polarized fermions (MCTDHF). Our approach is based on a mapping for operators in Fock space that allows a compact and efficient application of the Hamiltonian and solution of the MCTDHF equations of motion. Our implementation extends, builds on and exploits the recursive implementation of MCTDH for bosons (R-MCTDHB) package. Together with R-MCTDHB, the present implementation of MCTDHF forms the MCTDH-X package. We benchmark the accuracy of the algorithm with the harmonic interaction model and a time-dependent generalization thereof. These models consider parabolically trapped particles that interact through a harmonic interaction potential. We demonstrate, that MCTDHF is capable of solving the time-dependent many-fermion Schrödinger equation to an in principle arbitrary degree of precision and can hence yield \textit{numerically exact} results even in the case of Hamiltonians with time-dependent one-body 
and two-body potentials. As an application we study the problem of two initially parabolically confined and charged fermions tunneling through a barrier to open space. We demonstrate the validity of a model proposed previously for the many-body tunneling to open space of bosonic particles with contact interactions [Proc. Natl. Acad. Sci. USA {\bf 109}, 13521-13525 (2012)]. The many-fermion tunneling can be built up from sequentially happening single-fermion tunneling processes. The characteristic momenta of these processes are determined by the chemical potentials of trapped subsystems of smaller particle numbers: the escaped fermions convert the different chemical potentials into kinetic energy.
Using the two-body correlation function, we present
a detailed picture of the sequentiality of the process and are able to
tell tunneling from over-the-barrier escape. 
\end{abstract}

\pacs {}

\maketitle

\section{Introduction} 

The time-dependent many-body Schrödinger equation for interacting fermions governs systems from many different fields ranging from electron dynamics in molecules \cite{RMP-Mol} in quantum or theoretical chemistry, over graphene \cite{RMP-Graph} and (fractional) quantum Hall states \cite{FerCMT} in condensed matter, to the physics of quantum computation \cite{FerQC}, quantum simulation \cite{FerQS} or quantum dots and mesoscopic structures and interactions thereof \cite{FerQD,Annika,Annika_QD}, to name but a few. A general approach to deal with the time-dependent many-body Schrödinger equation for interacting fermionic particles is hence of great and general interest, especially also in view of the recent experimental demonstration of deterministic production of few-fermion systems \cite{SelimDeterministic} and their detailed investigation \cite{SelimDW,SelimFewMany,ZurnPRL} in the context of ultracold atoms. 

However, the solution of Schrödinger's equation for many-body systems presents a formidable, in most cases not analytically tractable problem. The exceptions to this statement include the Lieb-Liniger \cite{lieb} and Tonks-Girardeau \cite{Gir} models for one-dimensional bosonic particles and the harmonic interaction model for fermionic or bosonic particles of any dimension \cite{Cohen:85,Yan:03,HIM:Fer}. To date, unfortunately, no way has been found to scrutinize these or other analytical models to obtain solutions for the time-dependent Schrödinger equation for a general problem setting. Numerical methods that solve the Schrödinger equation are thus needed. Such numerical approaches also face limitations as the Hilbert space of many-body systems is growing exponentially with the particle number. Among the first approaches to the time-dependent many-body problem were so-called mean-field methods which reduce the intractably large problem by transforming the time-dependent many-body Schrödinger equation (TDSE) to an effective one-body problem. This transformation can be done in different ways. One approach is to make a mean-field ansatz for the state of the system and to derive the equations of motion for it by employing a time-dependent variational principle \cite{TDVP,DFVP}: demanding the stationarity of the functional action with respect to small variations of the parameters of said mean-field ansatz and demanding the solution to obey constraints (like, for instance, normalization), yields the equations of motion for the parameters (the time-dependent orbital[s]) in the mean-field ansatz. In the case of distinguishable particles, one obtains the time-dependent Hartree-type or self-consistent field equations \cite{DFVP}. For indistinguishable bosons, these equations of motion are called time-dependent Gross-Pitaevskii equation \cite{TDGP,GPbook} for a single-orbital ansatz and time-dependent multi-orbital mean-field equations \cite{TDMF} for a multi-orbital ansatz. For fermions they are named time-dependent Hartree(
-Fock) \cite{TDHF}. All these equations have in common that they prescribe the time-evolution of a single ([anti-]symmetrized) product of one-particle states or, in short, of one configuration. They therefore cannot describe correlations between the particles and their dynamics adequately since that necessitates multiple configurations.  

In order to straightforwardly go beyond the mean-field approach one allows not only one but all possible time-dependent configurations to contribute to the ansatz. One arrives, after a similar variational derivation as for the mean-field theories, at the multiconfigurational time-dependent Hartree approach (MCTDH) \cite{MCTDH,MCTDH-Rev,invariance} for distinguishable particles, the MCTDH for bosons in the case of indistinguishable bosons \cite{alon,stre} (for a dedicated version for the double-well special case, see Ref.~\cite{Reinh} and for the multi-layer MCTDH for bosons generalization, see Ref.~\cite{pete}), or the MCTDH for fermions (MCTDHF) \cite{unified,haxton,Nest,Kato,Scrinzi,MCTDHF:He} in the case of indistinguishable fermions, which is also often refered to as multiconfigurational
time-dependent Hartree-Fock. The generalization to multiconfigurational ansätze and hence improved accuracy comes at a price: the number of (symmetrized) basis states increases exponentially with the number of particles considered. The improvement in accuracy, however, can be crucial in order to describe the ongoing quantum dynamics, since it incorporates the evolution of correlations between the particles or degrees of freedom in the system. Moreover, the ansätze of MCTDH, MCTDH for bosons, and MCTDHF form a formally complete set of the respective many-body Hilbert spaces. Therefore, these methods can in principle provide the \textit{exact} solution of the full TDSE when a convergence with respect to the number of variational parameters in the description is achieved. This has been demonstrated for the bosonic case in Refs.~\cite{exact,exact2,exact3} and the fermionic case in, for instance, Ref.~\cite{MCTDHF:He}. 

Herein, we report on the implementation of MCTDHF that relies on the equivalence of many-body Hamiltonians in Fock space to a mapping \cite{MAP} that helps us to optimally scrutinize the sparsity of the Hamiltonian in configuration space.

We demonstrate how the algorithm converges to the exact solutions of the TDSE for the case of spin-polarized fermions even for Hamiltonians with time-dependent one- and two-body potentials. We show the convergence and exactness of the algorithm, we test it with the exactly solvable harmonic interaction model (HIM) and its time-dependent generalization (TDHIM) \cite{exact,exact2}.
Thereby, we demonstrate its capability to describe both ground states as well as
many-body dynamics with a time-dependent Hamiltonian,
 including time-dependent one-body
potentials and interactions, accurately.

We investigate charged fermions tunneling to
open space from an initial parabolic confinement similar to recent experimental
realizations \cite{SelimDeterministic,ZurnPRL}.
We thereby verify a model for the many-boson tunneling to open space
\cite{tunbos} to also hold in the fermionic case. We prepare a system of $N=2$ charged, spin-polarized fermions in the ground state of a parabolic trap. By subsequently transforming the potential to an open configuration with a barrier, we allow the fermions to escape to open space by tunneling. We monitor the process with the time-evolution of the momentum distributions and the one-body and
two-body correlation functions.  

We implemented the solution of MCTDHF equations of motion extending and exploiting our previous recursive implementation of the MCTDH for bosons, R-MCTDHB package \cite{ultr}. The resulting software is now capable of solving the time-dependent many-body Schrödinger equation for general indistinguishable particles. We name the resulting software MCTDH-X, since it computes the dynamics of multi-configurational wavefunctions of X=F fermions and X=B bosons. The MCTDH-X software is distributed under a copy-left license through the website \cite{ultr} and provides numerically exact results for bosons -- as shown in Refs.~\cite{exact,exact2} -- and -- as shown below -- numerically exact results for fermions. 

Let us mention here, that the present study considers long-range interactions in partially large grids. The evaluation of the respective two-body operators relies crucially on the so called ``interaction-matrix evaluation via successive transforms'' algorithm introduced in Refs.~\cite{exact3,exact2}. Without this algorithm the present and many recent investigations of realistic solutions of the TDSE as for instance in Refs.~\cite{exact,exact2,IMP1,IMP2,IMP3} would have been impossible.

The structure of this paper is as follows: in Sec.~\ref{Theory}, we introduce
the TDSE, derive the MCTDHF-equations of motion, describe our
implementation of it and show its accuracy.
Section \ref{tunnel:sec} discusses charged fermion tunneling to open space dynamics and Sec.~\ref{concl} gives conclusions and outlook.

\section{Theory, Implementation, and Exactness} \label{Theory}

\subsection{Time-Dependent Schrödinger Equation and Hamiltonian}\label{sec:TDSE}
The problem we aim to solve with the MCTDHF is the time-dependent Schrödinger equation for $N$ interacting fermionic particles. The TDSE reads:
\begin{equation}
 i \partial_t \vert \Psi \rangle = \hat{H} \vert \Psi \rangle. \label{TDSE}
\end{equation}
Here, the wavefunction $\vert \Psi \rangle$ and the Hamiltonian $\hat{H}$ depend on all the particle coordinates $\vec{r}_1,...,\vec{r}_N$ and time $t$. We investigate systems with at most two-body operators in the Hamiltonian:
\begin{equation}
 \hat{H} = \sum_k \hat{h}(\vec{r}_k;t) + \sum_{j<k} \hat{W} (\vec{r}_j,\vec{r}_k,t). \label{H1Q}
\end{equation}
Here, $\hat{h}(\vec{r}) = \frac{1}{2} \partial^2_{\vec{r}} + V(\vec{r},t)$, is the one-body Hamiltonian and $ \hat{W} (\vec{r},\vec{r}',t)$ represents the two-body interactions. Both terms are generally time-dependent. In this work, we employ dimensionless units throughout: the dimensionless Hamiltonian, Eq.~\eqref{H1Q}, is obtained by dividing the dimension-full Hamiltonian by $\frac{\hbar^2}{mL^2}$ ($m$ is the mass of the considered Fermions and $L$ is a conveniently chosen length scale).  

In second quantization, one uses field operators to represent the problem:
\begin{equation}
 \hat{\mathbf{\Psi}}(\vr,t) = \sum_k \ak(t) \phi_k(\vr;t).  \label{FO}
\end{equation}
The functions $\phi_k(\vr,t)$ are an orthonormal set of single-particle states or orbitals that build up a fully anti-symmetrized basis of the $N$-fermion Hilbert space.
Consequently, the Hamiltonian of Eq.~\eqref{H1Q} reads 
\begin{equation}
\hat{H} = \sum_{kq} h_{kq} (t) \adk \aq + \frac{1}{2} \sum_{kslq} W_{ksql}(t) \adk \ads \al \aq. \label{H2Q} 
\end{equation}
Here, we used the following notations for the one-body and two-body matrix elements $h_{kq}$ and $W_{ksql}$, respectively:
\begin{eqnarray}
 h_{kq}(t) &=& \int d\vr \phi^*_k (\vr,t) \hat{h}(\vr,t) \phi_q(\vr,t), \label{ME} \\ 
 W_{ksql}(t) &=& \iint d\vr' d\vr \phi^*_k (\vr',t) \phi^*_s (\vr,t) \hat{W}(\vr',\vr,t) \phi_q (\vr,t)\phi_l (\vr',t). \nonumber
\end{eqnarray}
In the following we will make use of the matrix elements of the reduced one- and two-body density matrices,
\begin{equation}
 \rho_{kq}(t) = \langle \Psi \vert \adk \aq \vert \Psi \rangle, \;\; \quad \rho_{kslq}(t) = \langle \Psi \vert \adk \ads \al \aq \vert \Psi \rangle, \label{RE}
\end{equation}
as well as the reduced one-body and two-body densities \cite{RDMs,RJG}
\begin{eqnarray}
 \rho^{(1)}(\vr,\vr',t) &=& \sum_{kq} \rho_{kq} \phi^*_k(\vr',t) \phi_q(\vr,t),\; \text{and}\\ \nonumber
 \rho^{(2)}(\vr_1,\vr_2,\vr'_1,\vr'_2,t) &=& \sum_{kslq} \rho_{kslq} \phi^*_k(\vr'_1,t) \phi^*_s(\vr_1,t)\phi_l(\vr'_2,t) \phi_q(\vr_2,t). \label{RDMs}
\end{eqnarray}
Now, all prerequisites for the derivation of the MCTDHF equations of motion are defined and we can proceed by doing so.

\subsection{MCTDHF equations of motion}
To derive the equations of motion of MCTDHF, we first formulate a general multiconfigurational ansatz for the wavefunction. Then, we use the time-dependent variational principle \cite{TDVP} to minimize the error in describing the solution of the TDSE with said ansatz.
The ansatz is obtained by truncating the field operator in Eq.~\eqref{FO} from an infinite to a finite sum of $M$ operators $\lbrace \ak (t) \rbrace_{k=1}^M$. Consequently, some expressions in the previous section \ref{sec:TDSE} become finite sums [cf. Eqs.~\eqref{FO},\eqref{H2Q},\eqref{RDMs}] or have a finite set of indices [cf. Eqs.~\eqref{ME},\eqref{RE}].

The wavefunction corresponding to a field operator in a finite set of $M$ time-dependent orbitals [see Eq.~\eqref{FO}] is the ansatz for MCTDHF and reads
\begin{equation}
 \vert \Psi \rangle = \sum_{\vec{n}} \Cn  \vert \vec{n};t \rangle = \sum_{\vec{n}} \Cn \prod_{i=1}^M \left( \adk(t) \right)^{n_k} \vert vac \rangle. \label{ansatz}
\end{equation}
Here, a vector notation $\vec{n}=(n_1,...,n_M)$ for the occupation numbers was invoked. In total, this ansatz contains $N_{conf} = \binom{M}{N}$ terms and as many time-dependent coefficients $\Cn$. The occupation number states or configurations $\lbrace \vert \vec{n};t \rangle \rbrace$ are fully anti-symmetrized products of the orbitals $\lbrace \phi_k (\vr;t) \rbrace_{k=1}^M$. The $N_{\text{conf}}$ coefficients and $M$ orbitals are the variational parameters in the derivation of the MCTDHF equations of motion.
The functional action of the TDSE [Eq.~\eqref{TDSE}] reads \cite{TDVP} as follows:  
\begin{equation}
\mathcal{S} = \int dt \left( \langle \Psi \vert \hat{H}- i\partial_t \vert 
\Psi \rangle+ \sum_{ij} \mu_{ij}(t)\left( \langle \phi_i \vert \phi_j 
\rangle - \delta_{ij} \right) \right). \label{action} 
\end{equation}
Here, the orthonormality of the orbitals $\lbrace \phi_k(x;t)
\rbrace_{k=1}^M$ is ensured by the Lagrange multipliers $\mu_{ij}(t)$ in
$\mathcal{S}$. From demanding the stationarity of this functional action with respect to variations of the coefficients $\Cn$ and the orbitals $\lbrace \phi_k(\vr,t) \rbrace_{k=1}^M$,
\begin{equation}
 \partial_{C^*_{\vec{n}}(t)} \mathcal{S}[\lbrace \Cn \rbrace, \lbrace \phi_k(\vr,t) \rbrace_{k=1}^M] \stackrel{!}{=}0 \;\; \forall \;{\vec{n}}; \qquad \partial_{\phi^*_k(\vr,t)} \mathcal{S} [\lbrace \Cn \rbrace, \lbrace \phi_k(\vr,t) \rbrace_{k=1}^M] \stackrel{!}{=} 0 \;\; \forall\; k,
\end{equation}
the equations of motion for the orbitals and the coefficients of the MCTDHF are obtained. To simplify the resulting equations of motion, and without loss of generality we use an invariance property of the ansatz [Eq.~\eqref{ansatz}], see Refs.~\cite{invariance,MCTDH}, and set $\langle \phi_k \vert i \partial_t \vert \phi_q \rangle =0$. For details of the derivation, see for instance Refs.~\cite{haxton,unified,MCTDHF:He}.
The obtained coefficients' equation of motion read
\begin{equation} 
 \mathcal{H}(t) \mathcal{C}(t) = i \partial_t \mathcal{C}(t);\qquad 
 \mathcal{H}_{\vec{m}\vec{m}'}(t) = \langle \vec{m};t \vert \hat{H} \vert \vec{m}' ; t \rangle. \label{CEOM}
\end{equation}
Here, $\mathcal{C}(t)$ collects all coefficients $C_{\vec{n}}(t)$ in a vector. The indexing of this vector is a key part of the implementation of MCTDHF and is described in the following subsection.
The orbitals' equation of motion read
\begin{eqnarray}
  i \partial_t \phi_j(\vr,t) &=& \mathbf{\hat{P}} \left. \Bigg(\hat{h} 
\phi_j(\vr,t) + \sum_{k,s,q,l=1}^M \lbrace \rho(t) \rbrace^{-1}_{jk}  
\rho_{kslq}(t) \hat{W}_{sl}(\vr,t) \phi_q(\vr,t) \right. \Bigg), \label{OEOM} \\  
 \mathbf{\hat{P}} &=& \mathbf{1} - \sum_{j'=1}^M \vert \phi_{j'} \rangle 
 \langle \phi_{j'} \vert. \nonumber
\end{eqnarray}
The projector $\mathbf{\hat{P}}$ emerges in the derivation from the Lagrange multipliers introduced in the functional action, Eq.~\eqref{action}, to ensure the orthonormality of the orbitals. The local time-dependent potentials $\hat{W}_{sl} (\vr;t) = \int d\vr' \phi^*_s(\vr',t) \hat{W}(\vr,\vr',t) \phi_l(\vr',t)$ were defined. Equations \eqref{CEOM} and \eqref{OEOM} form the core of the MCTDHF. 
The number of coefficients' equations [Eq.~\eqref{CEOM}] is $\binom{M}{N}$
and the number orbitals' equations [Eq.~\eqref{OEOM}] is $M$. In the latter
equation \eqref{OEOM}, the main computational effort is to compute $\mathcal{O}(M^4)$ two-body terms (containing $\rho_{kslq}$). With this scaling of the problem size, $\mathcal{O}(10)$ fermions are tractable at present without further optimization of the MCTDH-X software. 
We note, that the equations of motion, \eqref{CEOM} and \eqref{OEOM}, are of the same shape as the orbital equations of motion in the case of bosons \cite{unified}. The coefficients' and orbitals' equations form a coupled, integro-differential and generally non-linear set: The evaluation of Eq.~\eqref{CEOM} for the coefficients $\Cn$ necessitates the matrix elements $W_{ksql}(t)$ and $h_{kq}(t)$ [Eq.~\eqref{ME}] computed from the current set of orbitals $\lbrace \phi_k(\vr,t) \rbrace_{k=1}^M$. The propagation of Eq.~\eqref{OEOM} for the orbitals $\lbrace \phi_k(\vr,t) \rbrace_{k=1}^M$ necessitates the matrix elements $\rho_{kq}(t)$ and $\rho_{kslq}(t)$ [Eq.~\eqref{RE}] 
computed from the present set of coefficients $\Cn$. 
We move on and describe our implementation of MCTDHF.

\subsection{Hamiltonian as a mapping in Configuration Space}
Our MCTDHF implementation relies on a mapping for operators in Fock space as described in Ref.~\cite{MAP} and takes maximal advantage of the sparsity of the Hamiltonian in Fock basis [Eq.~\eqref{CEOM}].  
We use the address 
\begin{equation}
 I(h_1,...,h_{k_\mu})= 1 + \sum_{j=1}^{k_\mu} \binom{N + k_\mu - h_j}{ k_\mu + 1 - j } \label{indexing}
\end{equation}
to index all occupation number states of $N$ fermions in $M$ orbitals. Here $k_\mu$ is the number of holes, i.e., $0$ occupation numbers in the configuration $\vec{n}$ and $h_j$ the positions of these holes. Importantly, the indexing defined by Eq.~\eqref{indexing} allows us to write the coefficients $C_{\vec{n}}(t)$ in a compact vector notation $\mathcal{C}(t)$ as done in the MCTDHF coefficients equation of motion, Eq.~\eqref{CEOM}, above. We continue by illustrating how a general set of operators' actions can be cast in a compact and intuitive form by scrutinizing the indexing in Eq.~\eqref{indexing}.  The action of any of the operators that appear in the Hamiltonian Eq.~\eqref{H2Q} applied to a given configuration $\vert \vec{n}_1;t \rangle$ will yield a modified configuration $\alpha \vert \vec{n}_2 ; t \rangle$. Evidently, $\alpha$ depends on the occupation numbers $\vec{n}_1$ and on the applied creation and annihilation operators. Since all the configurations have an index assigned through Eq.~\eqref{indexing},
it is sufficient to know three numbers to apply \textit{any} operator: (i) the index $I_1$ of $\vert \vec{n}_1 ; t \rangle$, (ii) the index $I_2$ of $\vert \vec{n}_2 ;t \rangle$, and (iii) the prefactor $\alpha$.  
Let us consider the generic one-body operators $ \ad{4}\a{1}$ and $\ad{6} \a{3}$ and the configuration $\vert \vec{n}_1 ;t \rangle = \vert 1,1,1,0,1,0,0,1,0 \rangle$ as an example. One finds:
\begin{eqnarray*}
 \ad{4}\a{1} \vert 1,1,1,0,1,0,0,1,0 \rangle &=& \vert 0,1,1,1,1,0,0,1,0 \rangle; \\  \text{and} \qquad \ad{6} \a{3} \vert 1,1,1,0,1,0,0,1,0 \rangle &=& (-1) \vert 1,1,0,0,1,1,0,1,0 \rangle.
\end{eqnarray*}
In our example, we have $N=5,M=9$ and the number of holes is $k_\mu=4$.
Consequently, for $\ad{4} \a{1}$ with $I_1=I(4,6,7,9)=8$, we find $I_2=I(1,6,7,9)=73$ and $\alpha=1$ from Eq.~\eqref{indexing}. For $\ad{6} \a{3}$ with $I_1=8$, we find $I_2=I(3,4,7,9)=26$ and $\alpha=(-1)$. For many operators' actions on a configuration one finds that the respective prefactor $\alpha$ is zero and one does therefore not have to consider the action of the operator on that configuration. In order to minimize the effort in the evaluation of the MCTDHF coefficients equations of motion \eqref{CEOM}, we therefore only save the triples $I_1,I_2,\alpha$ when $\alpha\neq0$ in a dedicated custom data-type in our implementation of equation \eqref{CEOM}. The data-type is constructed with the following recipe: for every one- or two-body operator in the Hamiltonian \eqref{H2Q}, we analyze for every $I_1$ if $\alpha$ is zero. If so, we move on to the next configuration. Only if $\alpha$ is nonzero, $I_1,I_2$, and $\alpha$ are stored. The resulting set of triplets $(I_1,I_2,\alpha)$ for every operator in 
the Hamiltonian \eqref{H2Q}, for all configurations $\vert \vec{n}_1;t \rangle$, constitutes the most compact (and hence memory-efficient) way of applying the Hamiltonian to a given vector $\mathcal{C}(t)$ of coefficients. This allows for a faster evaluation of the coefficients equation of motion and for the handling of configuration spaces with a larger number of coefficients.    

\subsection{Benchmark with the time-dependent Harmonic Interaction model}
The HIM and TDHIM are models that are exactly solvable when a coordinate transform from Cartesian to center-of-mass and relative coordinates is applied to their Hamiltonians. Solutions are known for any spatial dimension and for bosonic and fermionic systems \cite{HIM:Fer}. 
The existence of exact solutions distinguishes the HIM models from other example problems such as, for instance, the Helium atom (cf. Ref.~\cite{MCTDHF:He}), for which approximate solutions with a very high accuracy are available -- but not exact ones.

Since the HIM and TDHIM Hamiltonians present a correlated many-body problem in Cartesian coordinates, these models are a good test for the accuracy of the MCTDHF algorithm which, of course, also works in Cartesian coordinates. 
Benchmarks have been performed previously in the case of the MCTDH for bosons with the HIM and its time-dependent generalization, the TDHIM in Refs.~\cite{exact,exact2}. 
Let us mention here, that the TDHIM presents a much tougher problem for the algorithm than typical physical problems like the tunneling to open space discussed in Sec.~\ref{tunnel:sec} because the TDHIM Hamiltonian has time-dependent both one-body and two-body terms.
In a study of the eigenstates of the HIM, we found that MCTDHF yields results with an arbitrarily large accuracy \cite{SupplMat}.

In this section, we asses the correctness of our implementation and the formal exactness of MCTDHF with the fermionic versions of the TDHIM. 

\begin{figure}
\includegraphics[width=0.9\textwidth,angle=-90]{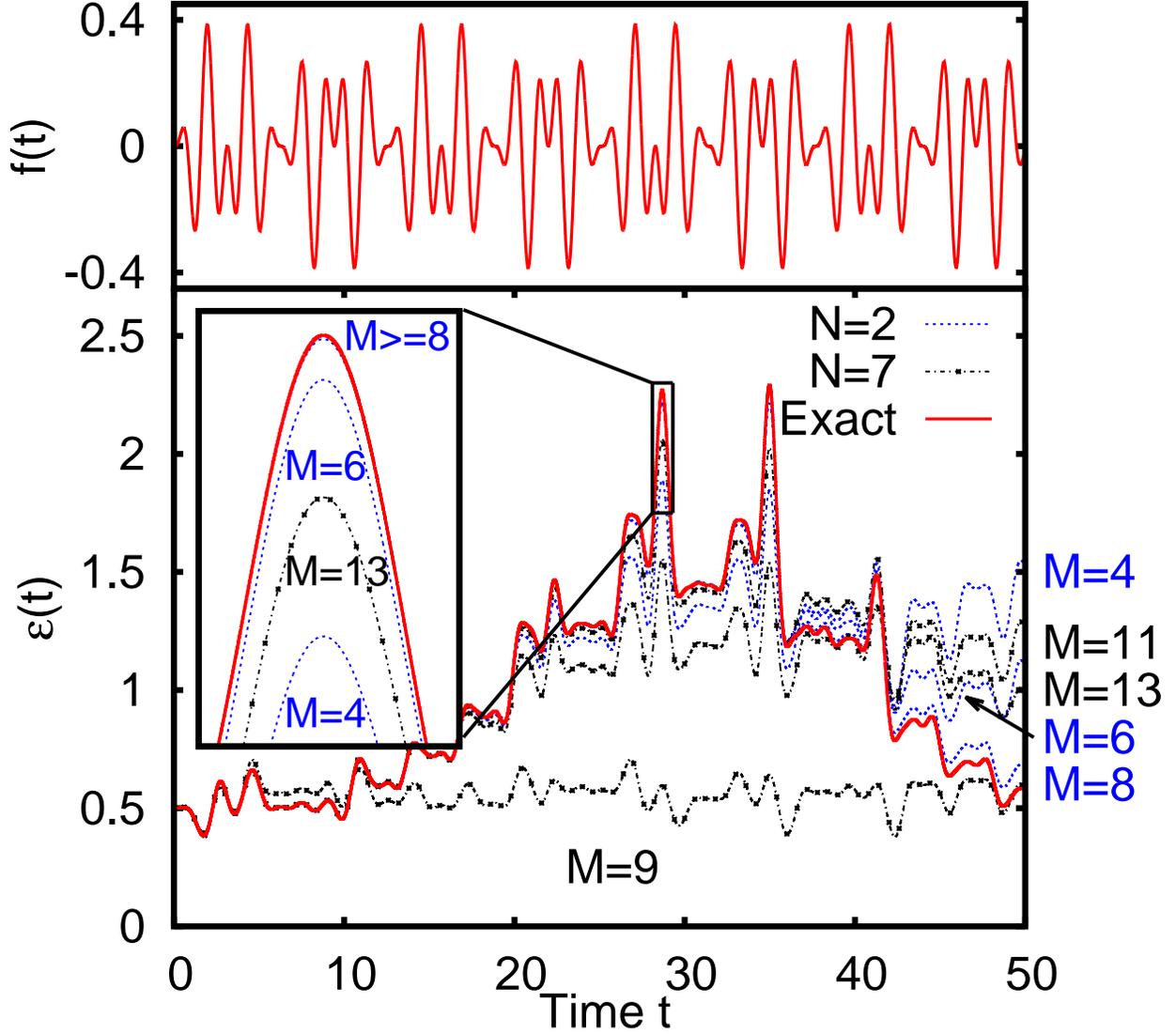}
\caption{Convergence of the MCTDHF for the TDHIM model for $N=2$ and $N=7$. The upper panel shows the time-dependent parameter $f(t)$ modulating both interactions and the frequency of the parabolical trap [cf. Eqs.~\eqref{TDHIM} and \eqref{TDHIMwK}]. The lower panel depicts the computed results for the time-dependent energy $\epsilon'_{\text{CM}}(t)$ for the exact solution (solid red line) and MCTDHF($M$) for increasing numbers of orbitals $M$ for two (blue dashed lines) and seven fermions (black dashed-dotted line with points). Fast convergence of the MCTDHF($M$) predictions for $\epsilon'_{\text{CM}}(t)$ to the exact value is observed for the present case of almost erratically time-dependent one-body and two-body parts in the Hamiltonian when $M$ is increased. We find that the MCTDHF prediction for $\epsilon'_{\text{CM}}(t)$ is indistinguishable from the exact energy for $M=10$ in the case of $N=2$ and for $M=15$ for the case of $N=7$. All quantities shown are dimensionless.}
\label{Fig:TDHIM}
\end{figure}

To arrive at a time-dependent generalization of the HIM, the TDHIM, we chose a Hamiltonian with time-dependent trapping frequency $\omega\equiv\omega_{TD}(t)$ and time-dependent interparticle interaction $K \equiv K_{TD}(t)$.
The obtained TDHIM Hamiltonian $\hat{H}'$ reads \cite{exact,exact2}:
\begin{equation}
 \hat{H}'(t)=\sum_{i=1}^N (- \frac{1}{2}\partial_{\vec{r}}^2 + \frac12 \omega_{TD}(t)^2 \vec{r}^2 ) +  K_{TD}(t) \sum_{i<j}^{j=N} \left( \vec{r}_i - \vec{r}_j \right)^2 \label{TDHIM}
\end{equation}
We now adopt the strategy in the Refs.~\cite{exact,exact2} and set
\begin{equation}
 \omega_{TD}(t) = \omega\left[ 1 + f(t) \right];\qquad K_{TD}(t) = K \left[ 1 - \frac{\omega_0^2}{2NK} f(t)\right]. \label{TDHIMwK}
\end{equation}
With this choice, we apply the coordinate transformations to relative $\vec{x}_j=\frac{1}{\sqrt{j(j+1)}} \sum_{i=1}^{j} (\vec{r}_{j+1} - \vec{r}_i), \qquad j=1,...,N-1$ and center-of-mass coordinates $\vec{x}_N=\sum_{i=1}^N \vec{r}_i $. We obtain
\begin{eqnarray}
 \hat{H}'_{rel}&=& \sum_{i=1}^{N-1} (- \frac12 \partial^2_{\vec{x}_i}+ \frac12 \delta^2_N \vec{x}^2_i);\qquad \delta_N=\sqrt{\omega^2 + 2NK}. \label{HIMTD} \\ 
\qquad \hat{H}'_{CM}(t)&=& - \frac12 \partial^2_{\vec{x}_N} + \frac12 \omega(t)^2 \vec{x}^2_N.
\end{eqnarray}
It's important to note that due to our choice of $\omega(t)$ and $K(t)$, the Hamiltonian of the relative problem is time-independent and identical to the one obtained for the HIM, i.e., when setting $f(t)\equiv 0$, see also \cite{SupplMat}. Moreover, the Hamiltonian of the center-of-mass problem $\hat{H}'_{CM}(t)$ defines the following albeit time-dependent, but \textit{one-body} problem that can be easily solved numerically exactly \cite{TDSE:1B}:
\begin{equation}
  i \partial_t \vert \Psi_{CM}(t) \rangle =\hat{H}'_{CM} (t)  \vert \Psi_{CM}(t) \rangle.
\end{equation}
The obtained time-dependent energy reads
\begin{eqnarray}
 E_{\text{TDHIM}}(t) &=& \epsilon_{rel} + \epsilon'_{CM}(t)  \\
\epsilon'_{CM} (t) &=& \langle \Psi \vert \hat{H}'_{CM} \vert \Psi \rangle. \nonumber
\end{eqnarray}
Since our choice for $K(t)$ and $\omega_{TD}(t)$ keeps $\delta_N=\sqrt{\omega^2 + 2NK}$ time-independent and identical to the $\delta_N$ in the case of the time-independent HIM Hamiltonian, $\epsilon_{rel}=D \sum_{i=1}^{N-1} \left[ \frac{2i+1}{2} \delta_{N} \right]$ remains also unchanged and time-independent (see \cite{SupplMat}). For computational convenience, we use the same parameters as in Refs.~\cite{exact,exact2}, $K_0 = 0.5$ and $\omega=1$ with $f(t)=\sin(t) \cos(2t) \sin(0.5t) \sin(2t)$. 
We plot the exact solution for $\epsilon'_{CM}(t)$ in Fig.~\ref{Fig:TDHIM} together with the MCTDHF predictions for $N=2$ and $N=7$ fermions for various numbers of orbitals $M$.

We find that the MCTDHF($M$) prediction converges to the exact result $\epsilon'_{CM}(t)$ rapidly for increasing $M$. This is analogous to the convergence of MCTDHB($M$) for the TDHIM model as shown in Refs.~\cite{exact,exact2}. We note here, that in the cases that the energies obtained from MCTDHF($M$) are identical to the exact one, a further increase of $M$ does not change the predictions of the method anymore: the wavefunction's time-evolution converges and represents a \textit{numerically exact} solution to the time-dependent Schr\"odinger equation. Importantly, the occupations of the least occupied orbitals remain negligibly small in the case of a converged solution of the TDSE with MCTDHF($M$). This renders a practical criterion for the convergence and numerical exactness of MCTDHF($M$) for any given application.

\section{Tunneling to open space of few-fermion systems}\label{tunnel:sec}
\begin{figure}
\includegraphics[width=0.9\textwidth, angle=-90]{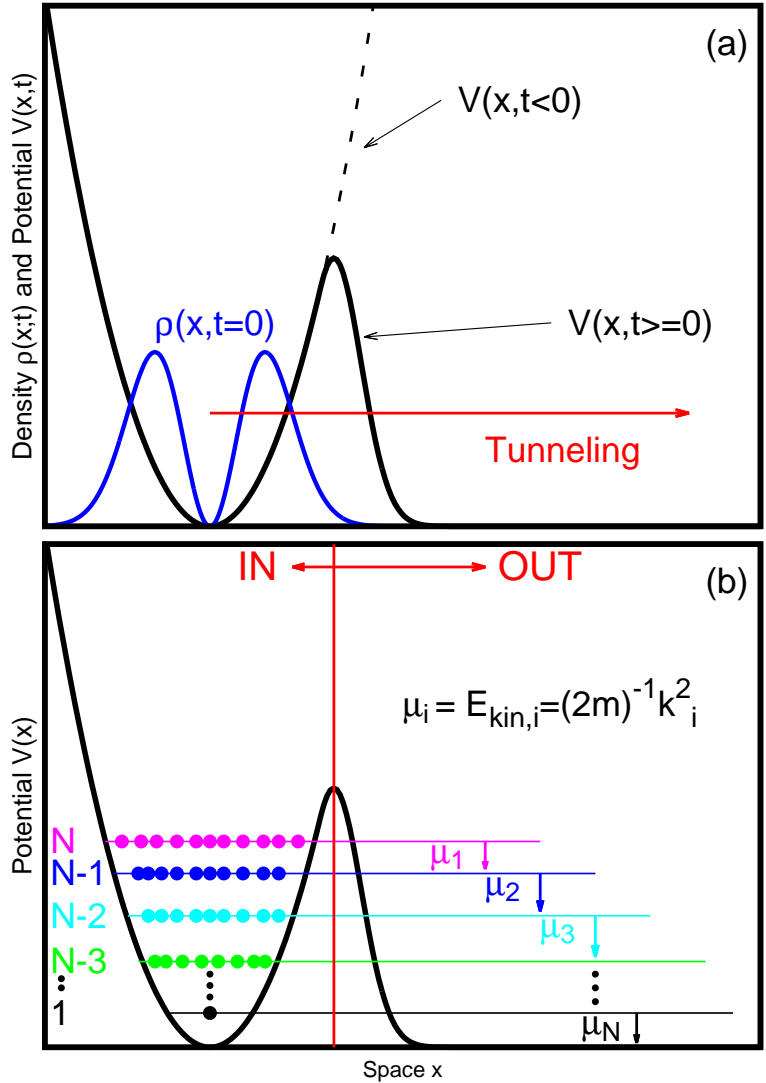}
\caption{Scheme and model for the fermion tunneling to open space process. {\bf (a)} shows the setup used to study how initially parabolically confined (black dashed line) fermions tunnel to open space when the potential is transformed to its open form (solid black line). For reference, an initial density is sketched (blue line). After the transformation of the potential, the system becomes unbound and tunnels through the barrier to open space. {\bf (b)} Model for the tunneling process. The one-dimensional Hilbert space is split into ``IN'' and ``OUT'' regions at the maximum of the barrier (vertical red line). The many-fermion process can be built up from concurrently happening single-fermion tunneling processes: each fermion escapes into ``OUT'' with a characteristic momentum. This characteristic momentum is in turn defined by the chemical potentials $\mu_i$ of the harmonically trapped, interacting system trapped in the ``IN'' region before the ejection. Neglecting interactions in ``OUT'', one 
obtains momenta $k_i= \sqrt{2m \mu_i}$  (Figure adapted from Refs.\cite{exact2,tunbos}, all quantities shown are in dimensionless units).} 
 \label{Fig:TunMod}
\end{figure}

There is outstanding experimental progress in the deterministic production of few-fermion systems \cite{SelimDeterministic} and detailed investigations on their properties \cite{SelimDW,SelimFewMany,ZurnPRL}. Of particular interest is here, that these experiments with few-fermion systems consider situations where one or several of the particles are escaping from an initial confinement to open space. 
The theoretical work entailing the above experiments deals mostly with the issue of the decay rates with respect to the possible different decay channels for fermions with internal structure, see for instance \cite{Blume,Rontani}. The correlation functions and their evolution have, however, not been systematically investigated yet.
This motivates us to apply MCTDHF to the problem of initially parabolically confined fermions that are allowed to tunnel through a potential barrier to open space. 
We stress here that our simulations below consider a system with a similar
one-body potential as the few-fermion systems in the experiments of the Heidelberg group \cite{SelimDeterministic,ZurnPRL}, but, both, the two-body potential and the constituents of the system differ from the experimental realization: in our simulations Coulomb and not contact interactions are considered and the constituents we consider are spin-polarized fermions and not fermions with an internal degree of freedom.
We adopt our scheme to model the process from previous work on bosons tunneling to open space \cite{tunbos:old,tunbos,exact2}: The system is initialized in the interacting groundstate of a parabolic potential (time $t<0$). Subsequently, at time $t\geq0$, the potential is transformed to an open form with a barrier that allows for the system to tunnel to open space.
For the sake of simplicity and convenience of computations, we use the same potential $V(\vec{r},t)$ without a threshold as Refs.~\cite{tunbos:old,tunbos,exact2}. See Fig.~\ref{Fig:TunMod}(a) for a plot. As interparticle interaction $\hat{W}(\vec{r},\vec{r}')$ we use the regularized Coulomb interaction from Ref.~\cite{Annika},
\begin{equation}
 \hat{W}(\vec{r},\vec{r}') = \frac{ \lambda_0 }{ \sqrt{ \vert \vec{r} - \vec{r}' \vert ^2  + \alpha^2 e^{ -\beta \vert \vec{r} - \vec{r}' \vert }}},
\end{equation}
where we set $\alpha=0.1$ and $\beta=100$, also as in Ref.~\cite{Annika}. 

Since the following considerations are for one spatial dimension, we are going to use the labels $x$ and $k$ synonymously for the vectors $\vec{r}$ and $\vec{k}$, respectively.
We restrict the present study to the case of $N=2$ fermions because the potential $V(x,t>0)$'s barrier (cf. Fig.~\ref{Fig:TunMod}(a)) is too small for larger particle numbers to still speak of a tunneling process: For the case of $N=3$ the energy per particle is comparable to the height of the barrier. Let us emphasize that we checked the results in the following for their convergence with respect to the number of orbitals $M$ in our computations and found that $M=6$ orbitals yield an essentially exact description. 

In the following, we study the dynamics for $N=2$ fermions for the case $\lambda_0=1.0$ and $\lambda_0=0.5$ and investigate if the model for the many-boson tunneling process in Ref.~\cite{tunbos,tunbos2} is predictive also for fermions. This model decomposes the one-dimensional Hilbert space into an ``IN'' and an ``OUT'' region and considers single-particle ejection processes from the viewpoint of energies. When a particle is ejected from ``IN'' to ``OUT'', the subsystem which is left behind remains with an energy $E_{IN}(N-1) = E_{IN}(N) - \mu_1$. Here, $\mu_1$ is the first chemical potential, see Fig.~\ref{Fig:TunMod},
$E_{IN}(N)$ is the numerical result of the two-fermion system and $E_{IN}(N-1)$ is the analytical result for the single-fermion ground state of the harmonic oscillator, $E_{IN}(1)=0.5$, since $N$ is $2$ . The emitted particle in ``OUT'' has the energy $\mu_1$ at its disposal which it converts to kinetic energy. Since the potential $V(x,t>0)$ is zero in almost all of ``OUT'', the emitted particle can approximately be considered as free and its kinetic energy is therefore given by $\mu_1 = \frac{k_1^2}{2 m}$. Consequently, a characteristic momentum $k_1=\sqrt{2 m \mu_1}$ will manifest in the momentum 
distribution $\rho(k,t)= \sum_{jq} \rho_{jq} \tilde{\phi}^*_j(k;t) \tilde{\phi}_q(k;t)$. Here, $\tilde{\phi}_q(x;t)$ denotes the Fourier transform of the orbital $\phi_q(k;t)$. The second emitted particle has the second chemical potential $\mu_2=E_{IN}(N-1)-E_{IN}(N-2)$, with $E_{IN}(N-2)=0$, to convert it to kinetic energy upon its ejection. This results in a second characteristic momentum $k_2=\sqrt{2m\mu_2}$ in the momentum distribution $\rho(k,t)$. By continuously applying this idea, the $N$-body tunneling process can be reconstructed by concurrently happening single-particle tunneling processes, cf. Fig.~\ref{Fig:TunMod}. 

\begin{figure}
\includegraphics[height=\textwidth, angle=-90]{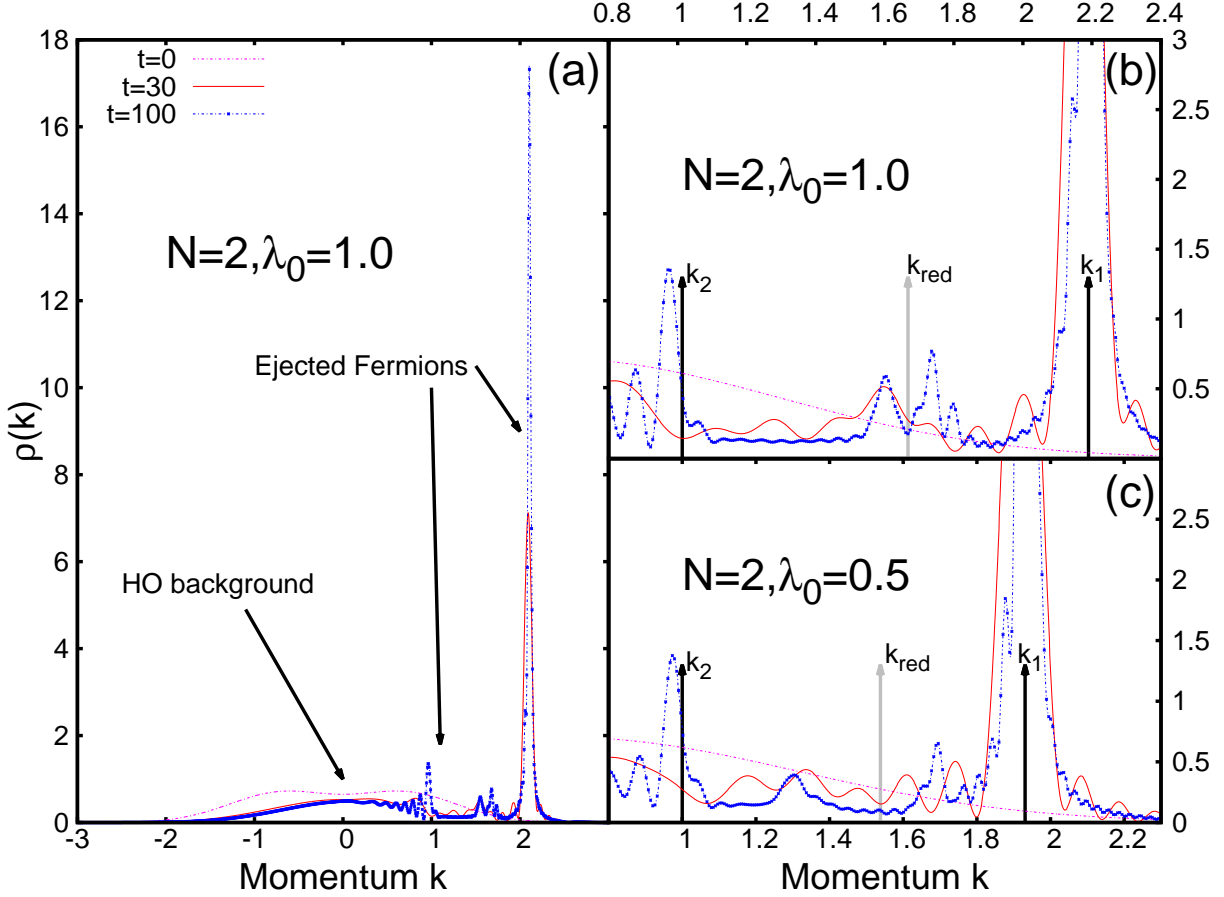}
\caption{Comparison of characteristic momenta of two-fermion tunneling to open space to model predictions. Panel (a) shows the overall momentum distributions $\rho(k)$ at times $t=0,30,100$ in the time-evolution of the tunneling process for $N=2$ and $\lambda_0=1.0$. The particles that escape to open space correspond to the peaks which emerge in the momentum distributions with time. These peaks' momentum distributions are enlarged for $\lambda_0=1.0$ in panel (b) and for $\lambda_0=0.5$ in panel (c).
According to the model in Fig.~\ref{Fig:TunMod}, the characteristic fermion ejection momenta are defined by the chemical potentials $\mu_i$ of the harmonically trapped, interacting system. The obtained momenta $k_i= \sqrt{2m \mu_i}$ are shown as black arrows in panels (b) and (c). The position of the main peaks agrees very well with the prediction of the model, $k_1,k_2$. The additional structure in the case of the stronger interactions in panel (b) can be attributed to a correlated two-body escape process (see text for further discussion). The momentum associated with this process is $k_{\text{red}}= \sqrt{2 \kappa (\mu_1+\mu_2)}=1.613$ where $\kappa=\frac{1}{2}$ is the reduced mass of the two-fermion system [see gray arrow in panel (b)]. For the weaker interactions $\lambda_0=0.5$ in panel (c), the two-fermion tunneling or escape seems to be less favorable, since no clear features are seen at the respective momentum $k_{\text{red}}=1.538$. All quantities shown are dimensionless. See text for further discussion.} 
 \label{Fig:Tun-ks}
\end{figure}

The momenta obtained from the model for the $N=2$ and $\lambda_0=1.0$ case are $k_1=2.103$ and $k_2=1.0$. For the $N=2$, $\lambda_0=0.5$ case, we find $k_1=1.931$ and $k_2=1.0$. We plot the exact time-evolution of the momentum density $\rho(k,t)$ in Fig.~\ref{Fig:Tun-ks}. We find that the model for the tunneling process -- albeit originally devised for bosonic particles with a contact interparticle interaction potential -- predicts the momenta of the fermions with Coulomb interactions emitted to open space with a remarkable degree of accuracy, see Fig.~\ref{Fig:Tun-ks}(b),(c). Upon closer inspection of Fig.~\ref{Fig:Tun-ks}(b), we find a peak structure in the momentum distribution in between $k_1=2.103$ and $k_2=1.0$. This structure corresponds to a kind of correlated pair ejection (see correlation functions $g^{(1)}$ and $g^{(2)}$ below) and the energy of the whole system $E_{IN}(N=2)=2.712$ is sufficiently over the height of the barrier for $\lambda_0=1.0$. Furthermore, one gets a good estimate of the momentum of the process, $k_{\text{red}}$, 
with the assumption that a particle with energy $\mu_1+\mu_2=E_{IN}(N=2)$ and the reduced mass $\kappa=\frac{1}{2}$ was ejected:
\begin{equation*}
 k_{\text{red}}= \sqrt{2 \kappa E_{IN}(N=2)}.
\end{equation*}
One finds $k_{\text{red}}=1.613$ or $\lambda_0=1$ and $k_{\text{red}}=1.538$ for $\lambda_0=0.5$. Since there is no clear feature around $k_{\text{red}}$ in the momentum distribution $\rho(k,t)$ in Fig.~\ref{Fig:Tun-ks}(c) for the case of $\lambda_0=0.5$, we infer that in that case the total energy of the system is not sufficiently over the barrier for the correlated two-body ejection to happen. Let us remark here that this is an interesting difference to the tunneling dynamics of the bosonic systems investigated in Refs.~\cite{tunbos,tunbos2,exact2}. We leave it as subject of further investigations to determine if the emergence of $k_{\text{red}}$ in the momentum distribution and how the feature is modified for a larger
number of fermions or bosons with the same Coulomb interactions as in the present study. 
We infer from the remarkable agreement of the predicted momenta and the momentum peaks in $\rho(k,t)$ in Fig.~\ref{Fig:Tun-ks} that the bulk of the many-fermion tunneling to open space process is -- like its many-boson counterpart -- built up by single particle processes with characteristic momenta $k_1,k_2,...$ determined by chemical potentials $\mu_1,\mu_2,...$ of trapped subsystems of decreasing particle number $N,N-1,\dots$ . 

We move on to investigate the dynamics of the correlation functions in momentum space $\vert g^{(1)}(k,k';t)\vert^2$ and in real space, $\vert g^{(1)}(x,x';t)\vert^2$.
The definition of $g^{(1)}$ is \cite{RJG}
\begin{equation*}
 g^{(1)}(\xi,\xi',t)= \frac{\rho^{(1)}(\xi,\xi',t)}{\sqrt{\rho^{(1)}(\xi,\xi,t) \rho^{(1)}(\xi',\xi',t)}},
\end{equation*}
where $\xi=k$ for the momentum space correlation function and $\xi=x$ for the real-space correlation function. The reduced one-body densities in momentum space, $\rho^{(1)}(k,k',t)$ are obtained from equation Eq.~\eqref{RDMs} by applying a Fourier transform to the orbitals.
In the case of uncorrelated particles characterized by the momenta $k$ and $k'$
or coordinates $x$ and $x'$, the correlation function is $1$, while it is $0$
for correlated particles. Its pattern allows to analyze whether the
particles tunnel sequentially or not (see in this context also the two-body correlation $g^{(2)}$ below). From the value of the correlation function $g^{(1)}$, it can further be inferred whether certain points in space $(x,x')$
or combinations of momenta $(k,k')$ correspond to the same ($\vert g^{(1)} \vert =1$) or to different ($\vert g^{(1)} \vert =0$) orbitals.

\begin{figure}
\centering
 \includegraphics[width=0.9\textwidth,angle=-90]{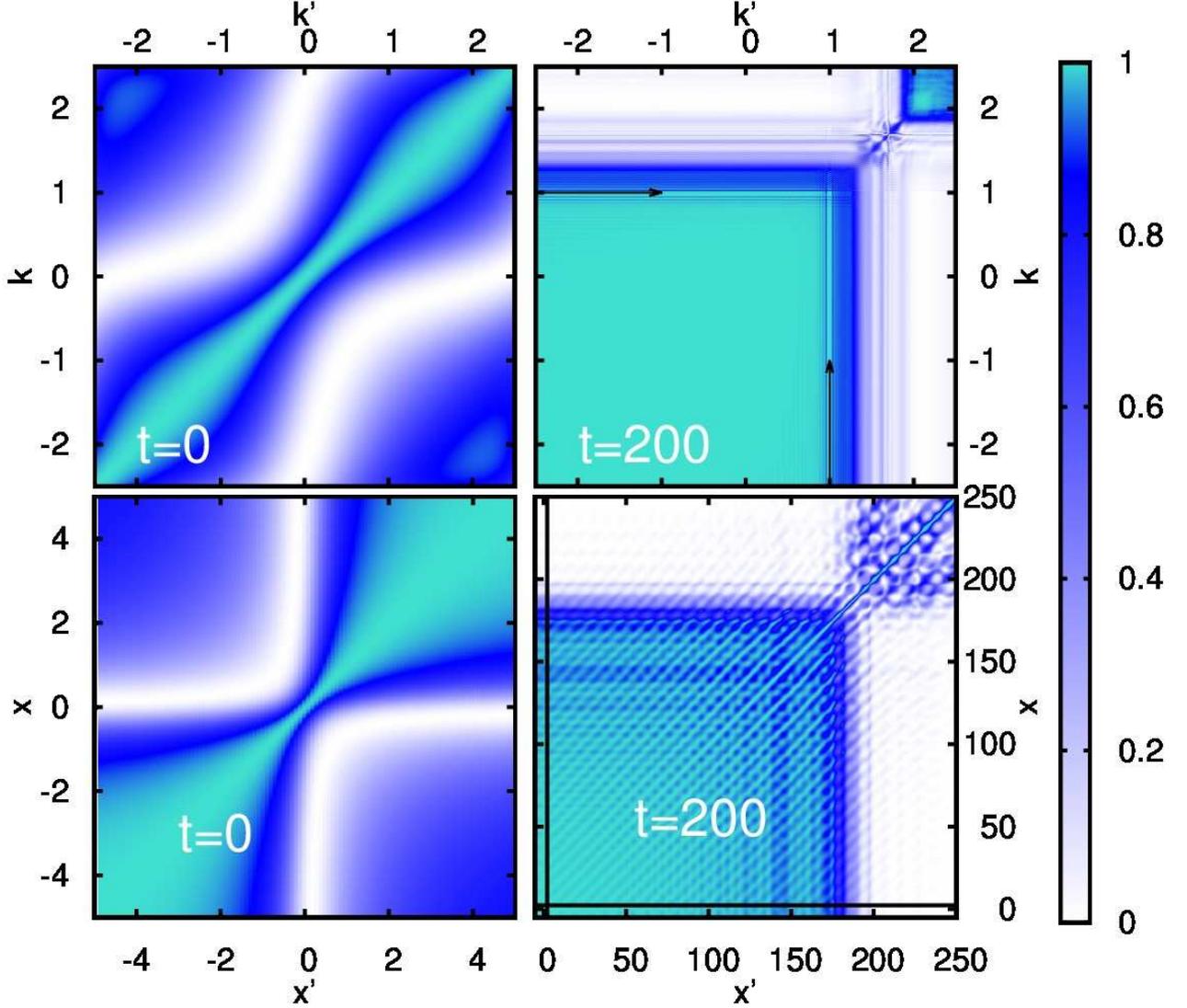}
 \caption{Spatial and momentum one-body correlation functions of fermion tunneling to open space. We plot the correlation functions $\vert g^{(1)}(x',x;t) \vert^2$ and $\vert g^{(1)}(k',k;t) \vert^2$ in real and momentum space in the top and bottom rows, respectively, for the $\lambda_0=1$ case.  In comparison to the bosonic case (see Refs.~\cite{tunbos,tunbos2}, we observe a generally more strongly correlated behavior in both real and momentum space for the ground states depicted in the left column. In the momentum correlations (top left and top right panels), a line structure emerges at the momenta corresponding to the escaping fermions, $k_1,k_2$ with time. While the fermions that still reside in the well at $k \approx k' \approx 0$ are almost uncorrelated ( $\vert g^{(1)} \vert^2 \approx 1$, the fermion escaping with momentum $k_1$ is uncorrelated with those at rest ( $\vert g^{(1)}(k'\approx k_1,k \approx 0;t=200) \vert^2 \approx 0$, see upper right panel). Interestingly, the fermion escaping with $k_2$ 
is correlated with the one at rest ( $\vert g^{(1)}(k'\approx k_2,k \approx 0;t=200) \vert^2 \approx 1$, see black arrows in top right panel). However, the $k_2$ velocity is embedded by thin lines where $\vert g^{(1)}(k'=k_2\pm\epsilon,k;t=200)\vert^2<1$ (see dark lines next to the arrows in the top right panel). The periodic structure in the real-space correlation in the bottom right panel reflects the momentum correlations in the top right panel. To guide the eye, we mark the top of the barrier by the black lines. The difference in the escape velocities $k_1,k_2$ leads to a strong spatial correlation (see rectangular areas with $\vert g^{(1)}(x,x';t=200) \vert^2 \approx 0$ on the off-diagonal of the bottom right panel). All quantities shown are dimensionless, see text for further discussion.}
 \label{Fig:Tun-gs}
\end{figure}

As a first observation (left column of Fig.~\ref{Fig:Tun-gs}), we find that the correlation of the initial state of the fermionic systems is much stronger than in the bosonic counterpart investigated in Ref.~\cite{tunbos,exact2}, which is essentially uncorrelated, i.e. $\vert g^{(1)}\vert^2 \approx 1$. Throughout the tunneling process, we find an emergence of a line structure of correlations in $\vert g^{(1)}(k,k';t)\vert^2$: around the momenta $k_1,k_2$ at which particles are ejected from the well to open space, see top right panel of Fig.~\ref{Fig:Tun-gs}. We deduce from this line structure that the tunneling process can be controlled by adjusting the potential in the outer region in analogy to the bosonic tunneling to open space process \cite{tunbos,tunbos2}. Upon detailed comparison of the emergent line structure with respect to the bosonic case \cite{tunbos} in which the lines are simply marked by $\vert g^{(1)} \vert\rightarrow 0$, the following interesting difference is found: at the second momentum 
$k_2$ one actually finds $\vert g^{(1)}(k_2,k)\vert \approx 1$ which is enclosed by lines at a small distance $\iota$ with $\vert g^{(1)}(k_2\pm\iota,k)\vert < 1$ (cf. black arrows and their vicinity in top right panel of Fig.~\ref{Fig:Tun-gs}). The different correlation properties of $k_1$ and $k_2$ hint that the fermions traveling at these velocities are sitting in different orbitals. 
This indicates that the two fermions escape sequentially. For an investigation of the reasons for this difference to the case of the bosonic tunneling to open space, where the single-particle processes happen concurrently, see the discussion of the two-body correlations below. 

The momenta $k_1,k_2$ are reflected in $\vert g^{(1)}(x,x';t)\vert^2$ as periodic structures. The large difference between $k_1$ and $k_2$ causes one of the fermions to escape much faster than the other. This leads to a strong correlation, i.e., $\vert g^{(1)}(x,x';t)\vert^2 \approx 0 $ on the off-diagonal (cf. ``white rectangles'' on the off-diagonal in the bottom right panel of Fig.~\ref{Fig:Tun-gs}). Moreover, this strong correlation on the off-diagonal of $g^{(1)}$ can be interpreted as a feature of the process being sequential. 

\begin{figure}
\centering
\includegraphics[width=\textwidth]{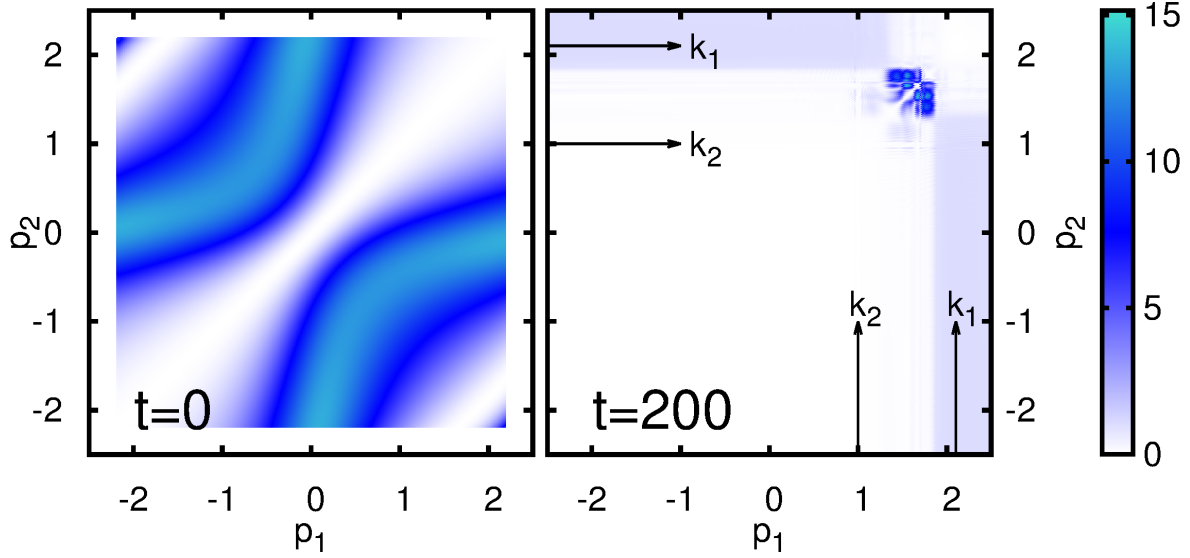}
\caption{Momentum two-body correlation functions of fermion tunneling to open
space. We plot the correlation functions $g^{(2)}(p_1,p_2,p_1,p_2;t)$ in momentum
space, for the $\lambda_0=1$ case for times $t=0$ (left) and $t=200$ (right).
In both panels, the Pauli exclusion principle is manifest, since
$g^{(2)}(p,p,p,p)=0$. For the initial state in the left panel, we find large
values of $g^{(2)}(p_1,p_2,p_1,p_2)$ where $\left| g^{(1)}(p_1,p_2) \right|^2$
is small and vice versa (cf. top left panel of Fig.~\ref{Fig:Tun-gs}). This
also vaguely holds for the $t=200$ plot (cf. top right panel of
Fig.~\ref{Fig:Tun-gs}). In the right panel at $t=200$, the black arrows mark the
escape momenta obtained from our model and the momentum distributions
(cf. Fig.~\ref{Fig:Tun-ks}). The $p_{1/2}=k_2 = 1$ tunneling at
$p_{1/2} = k_2 = 1.0$ is anti-correlated or
anti-bunched ($g^{(2)}<1$), whereas the tunneling at $k=2.103$ is correlated or bunched ($g^{(2)}>1$). The found correlated pair tunneling process emerges as a strongly bunched structure on the diagonal around $k_{\text{red}}=1.613$, i.e., $g^{(2)}(k_{\text{red}},k_{\text{red}},k_{\text{red}},k_{\text{red}})\gg1$. 
 All quantities shown are dimensionless, see text for further discussion.
}
 \label{Fig:Tun-g2s}
\end{figure}

Since the overall features in the momentum distributions (peaks corresponding to chemical potentials) and the one-body correlation functions (line structure in $k$-space) are quite similar in the case of fermions and bosons, we conclude that (i) the model in Fig.~\ref{Fig:TunMod} initially put forward for bosons is indeed also predictive for the case of fermions and (ii) the process for fermions is likely to allow for a control by the threshold value of the potential $T=\lim_{x\rightarrow \infty} V(x,t>0)$ in the same way as it does for bosons \cite{tunbos2}:
by changing the threshold $T$, the peaks in the momentum distribution can be
shifted and are turned off when $T$ becomes larger than the chemical
potential corresponding to the respective process.
The newly discovered feature of correlated two-body tunneling as well as the found differences in the time-evolution of the one-body correlation function are motivating for a more detailed investigation of the process using the two-body correlations. These can be quantified by the two-body correlation function 
\begin{eqnarray*}
 g^{(2)}(\xi_1,\xi_2,\xi'_1,\xi'_2,t) = \frac{\rho^{(2)}(\xi_1,\xi_2,\xi'_1,\xi'_2,t)}{\sqrt{\rho^{(1)}(\xi_1,\xi_1,t) \rho^{(1)}(\xi_2,\xi_2,t)\rho^{(1)}(\xi'_1,\xi'_1,t)\rho^{(1)}(\xi'_2,\xi'_2,t)}}.
\end{eqnarray*}
Here, $\rho^{(2)}$ is the reduced two-body density (cf. Eq.~\eqref{RDMs}). In Fig.~\ref{Fig:Tun-g2s}, we depict a plot of $g^{(2)}$ in momentum space ($\xi=p$) for the case of $\lambda_0=1.0$. As a first observation, we find that the two-body correlation function in momentum space exhibits a structure which is complementary to the one-body correlation function $g^{(1)}$ in momentum space (top row of Fig.~\ref{Fig:Tun-gs}): whereever $g^{(1)}(p_1,p_2)$ is comparatively large, $g^{(2)}(p_1,p_2,p_1,p_2)$ is comparatively small and vice versa. Moreover, we can decipher the details of the mechanism of the two-body tunneling to open space: a particle escaping with $p_1=k_1=2.103$ is bunched with the other particle not having the same momentum, i.e., $g^{(2)}(k_1,a,k_1,a)>1$ only for $a\neq k_1$ and $0$ otherwise (see arrows at $k_1=2.103$ in Fig.~\ref{Fig:Tun-g2s}). This means that it is likely that the second particle travels at a different velocity if the first one is found traveling at $k_1$. Moreover, a particle traveling at velocity $p_1=k_2=1.0$ is bunched with the other particle having the momentum $k_2$, i.e., $g^{(2)}(k_2,a,k_2,a)>1$ only for $a\approx k_1$ and $0$ otherwise. This underlines the sequential nature of the process: the second particle at $k_2$ can start its escape only,
once the other particle has already escaped with the momentum $k_1$ (see arrows at $k_2=1.0$ in Fig.~\ref{Fig:Tun-g2s}).
Moreover, the two-body correlation function demonstrates clearly that the peaks in the momentum distributions at $k_{\text{red}}$ (cf. top right panel of Fig.~\ref{Fig:Tun-ks}) indeed correspond to a correlated two-body escape: two particles escape with similar momenta as $g^{(2)}(p_1,p_2,p_1,p_2)$ exhibits strong bunching on the diagonal around $p_1\approx p_2 \approx k_{\text{red}}$. This bunching feature in $g^{(2)}$ on the diagonal disappears for the cases in which the peaks around $k_{\text{red}}$ are absent in the momentum distributions, i.e., for $\lambda_0=0.5$ and $\lambda_0=0$ (not shown). This corroborates the finding that the two-body escape we see is indeed an ``over-the-barrier'' effect and not tunneling.

\section{Conclusions and Outlook}\label{concl}
In the present work, we have demonstrated that the MCTDHF theory and algorithm are capable of solving general time-dependent many-fermion problems to an arbitrary degree of accuracy, i.e., we have shown the \textit{numerical exactness} of MCTDHF for the spin-polarized case. We found exponentially converging groundstate energy eigenvalues for the eigenstates of the harmonic interaction model. Solving the time-dependent harmonic interaction model with MCTDHF we demonstrated that even with erratically time-dependent one-body and two-body terms in the Hamiltonian, an exact solution of the time-dependent Schrödinger equation can still be obtained. 

We further studied the tunneling to open space process of few-fermion systems composed of charged spin-polarized particles. We assess the validity of a model put forward for the process for bosons with a contact interaction potential also in the present case of fermions interacting with a regularized Coulomb interaction. The prediction of the model for the escape momenta of the fermions is remarkably accurate. From a comparison of the momentum distributions and one-body correlation functions in the process with the model, we infer that the many-fermion tunneling to open space process is built up from sequentially happening single particle processes whose characteristic momenta $k_1,k_2,...$ emerge from the chemical potentials $\mu_1,\mu_2,...$ of trapped subsystems of decreasing particle number $N,N-1,...$. For the case, where the total energy of the system is sufficiently above the barrier height of the single-particle potential, we find an additional signature in the momentum 
distributions, which is associated with a correlated two-body escape process.
Using the two-body momentum correlation function, we were able to demonstrate the sequentiality of the process and that the discussed two-body process corresponds
to an ``over-the-barrier'' escape and not tunneling.

To scrutinize and investigate the emergent features in the correlation functions for a larger number of particles and other types of interactions is a subject of future work.

As further future applications of our MCTDHF implementation, we envisage for instance studies of statistical relaxation and chaos \cite{socc1,barnali}, quantum turbulence \cite{socc} and vortices \cite{vort} as well as Hubbard Hamiltonians \cite{prep}, electronic decay processes \cite{Annika}, and high harmonic generation \cite{Lewenstein94}.

\begin{acknowledgements}
Benchmark results made available by and helpful discussions with Hans-Dieter Meyer as well as the hospitality of Vanderlei S. Bagnato and Marios C. Tsatsos in the CePOF of the IFSC, Sa\~{o} Carlos, Sa\~{o} Paulo, Brazil are gratefully acknowledged. A.~U.~J.~L. gratefully acknowledges financial support by the Swiss SNF and the NCCR Quantum Science and Technology. E.~F. gratefully acknowledges funding from
the Research Council of Norway (RCN) through CoE Grant No. 179568/V30 (CTCC).
Helpful discussions with Christoph Bruder are acknowledged.
\end{acknowledgements}

\clearpage

\clearpage

\begin{thebibliography}{References}
\bibitem{Blume} S.~E. Gharashi and D. Blume, Tunneling dynamics of two interacting one-dimensional particles, Phys. Rev. A {\bf 92}, 033629 (2015).

\bibitem{Rontani} M. Rontani, Tunneling Theory of Two Interacting Atoms in a Trap, Phys. Rev. Lett. {\bf 108}, 115302 (2012).
\bibitem{RMP-Mol} E. Deumens, A. Diz, R. Longo, and Y. Öhrn, Time-dependent theoretical treatments of the dynamics of electrons and nuclei in molecular systems, Rev. Mod. Phys. {\bf 66}, 917 (1994); H. Feldmeier and J. Schnack, Molecular dynamics for fermions, Rev. Mod. Phys. {\bf 72}, 655 (2000). 

\bibitem{RMP-Graph} V.~N. Kotov, B. Uchoa, V.~M. Pereira, F. Guinea, and A.~H.~C. Neto, Electron-Electron Interactions in Graphene: Current Status and Perspectives, Rev. Mod. Phys. {\bf 84}, 1067 (2012).

\bibitem{FerCMT} R.~B. Laughlin, Quantized motion of three two-dimensional electrons in a strong magnetic field, Phys. Rev. B {\bf 27}, 3383 (1983); ibid. Anomalous Quantum Hall Effect: An Incompressible Quantum Fluid with Fractionally Charged Excitations, Phys. Rev. Lett. {\bf 50}, 1395 (1983).

\bibitem{FerQC} C. Nayak, S.~H. Simon, A. Stern, M. Freedman, and S. Das Sarma, Non-Abelian anyons and topological quantum computation, Rev. Mod. Phys. {\bf 80}, 1083 (2008).

\bibitem{FerQS} M. Köhl, H. Moritz, T. Stöferle, K. Günter, and T. Esslinger, Fermionic Atoms in a Three Dimensional Optical Lattice: Observing Fermi Surfaces, Dynamics, and Interactions, Phys. Rev. Lett. {\bf 94}, 080403 (2005); D. Greif, L. Tarruell, T. Uehlinger, R. Jördens, and T. Esslinger, Probing Nearest-Neighbor Correlations of Ultracold Fermions in an Optical Lattice, Phys. Rev. Lett. {\bf 106}, 145302 (2011).

\bibitem{FerQD} C. Kloeffel, P.~A. Dalgarno, B. Urbaszek, B.~D. Gerardot, D. Brunner, P. M. Petroff, D. Loss, and R. J. Warburton, Controlling the Interaction of Electron and Nuclear Spins in a Tunnel-Coupled Quantum Dot, Phys. Rev. Lett. {\bf 106}, 046802 (2011).
\bibitem{Annika_QD} F.~M. Pont, A. Bande, and L.~S. Cederbaum, Controlled energy-selected electron capture and release in double quantum dots, Phys. Rev. B {\bf 88}, 241304(R) (2013).
\bibitem{Annika} A. Bande, Electron dynamics of interatomic Coulombic decay in quantum dots induced by a laser field, J. Chem. Phys. {\bf 138}, 214104 (2013).

\bibitem{SelimDeterministic} F. Serwane, G. Zürn, T. Lompe, T.~B. Ottenstein, A.~N. Wenz and S. Jochim, Deterministic Preparation of a Tunable Few-Fermion System, Science {\bf 332}, 6027 (2011).
\bibitem{ZurnPRL} Pairing in Few-Fermion Systems with Attractive Interactions, G. Zürn, A. N. Wenz, S. Murmann, A. Bergschneider, T. Lompe, and S. Jochim, Phys. Rev. Lett. {\bf 111}, 175302 (2013).
    
\bibitem{SelimDW} S. Murmann, A. Bergschneider, V.~M. Klinkhamer, G. Zürn, T. Lompe, and S. Jochim, Two Fermions in a Double Well: Exploring a Fundamental Building Block of the Hubbard Model, Phys. Rev. Lett. {\bf 114}, 080402 (2015).
\bibitem{SelimFewMany} A.~N. Wenz, G. Zürn, S. Murmann, I. Brouzos, T. Lompe, S. Jochim, From Few to Many: Observing the Formation of a Fermi Sea One Atom at a Time, Science {\bf 342}, 457 (2013).

\bibitem{lieb} E.~H. Lieb and W. Liniger, Exact Analysis of an Interacting Bose Gas. I. The General Solution and the Ground State, Phys. Rev. {\bf{130}}, 1605 (1963).
\bibitem{Gir} M.~D. Girardeau, E.~M. Wright, and J.~M. Triscari, Ground-state properties of a one-dimensional system of hard-core bosons in a harmonic trap, Phys. Rev. A {\bf 63}, 033601 (2001).
\bibitem{HIM:Fer} M.~A. Za\l{}uska-Kotur, M. Gajda, A. Or\l{}owski, and J. Mostowski, Soluble model of many interacting quantum particles in a trap, Phys. Rev. A {\bf 61}, 033613 (2000).
\bibitem{Yan:03} J. Yan, Harmonic Interaction Model and Its Applications in Bose-Einstein Condensation, J. Stat. Phys. {\bf 113}, 623 (2003).
\bibitem{Cohen:85} L. Cohen and C. Lee, Exact reduced density matrices for a model problem, J. Math. Phys. {\bf 26}, 3105 (1985).

\bibitem{DFVP} J. Frenkel, P.~A.~M. Dirac, Wave mechanics, Oxford Univ. Press, Oxford 435 (1934); ibid. Proc. Cambridge Phil. Soc., \textbf{26}, 376 (1930).
\bibitem{TDVP} P. Kramer and M. Saraceno, {\it Geometry of the Time-Dependent Variational Principle in Quantum Mechanics}, Lecture Notes in Physics {\bf 140}, (Springer, Heidelberg, 1981).

\bibitem{TDGP} L.~P. Pitaevskii, Vortex Lines in an Imperfect Bose Gas, JETP {\bf 40}, 646 (1961); E.~P. Gross, Structure of a quantized vortex in boson systems, Il Nuovo Cimento {\bf 20}, 454 (1961).
\bibitem{GPbook} L. Pitaevskii and S. Stringari, {\it Bose-Einstein Condensation} (Oxford University Press, New York, 2003).
\bibitem{TDMF} O.~E. Alon, A.~I. Streltsov, and L.~S. Cederbaum, Time-dependent multi-orbital mean-field for fragmented Bose-Einstein condensates, Phys. Lett. A {\bf 362}, 453 (2007).
\bibitem{TDHF} A.~D. McLachlan and M.~A. Ball, Time-Dependent Hartree-Fock Theory for Molecules, Rev. Mod. Phys. {\bf 36}, 844 (1964).

\bibitem{MCTDH} H.-D. Meyer, U. Manthe, and L.~S. Cederbaum, The multi-configurational time-dependent Hartree approach, Chem. Phys. Lett. {\bf 165}, 73 (1990).
\bibitem{MCTDH-Rev} M.~H. Beck, A. Jäckle, G.~A. Worth, and H.~D. Meyer, The multiconfiguration time-dependent Hartree method: A highly efficient algorithm for propagating wavepackets, Phys. Rep. {\bf 324}, 1 (2000).
\bibitem{invariance} U. Manthe, H.-D. Meyer and L. S. Cederbaum, Wave-packet dynamics within the multiconfiguration Hartree framework: General aspects and application to NOCl, J. Chem. Phys. {\bf 97}, 3199 (1992).

\bibitem{alon} O.~E. Alon, A.~I. Streltsov, and L.~S. Cederbaum, Multiconfigurational time-dependent Hartree method for bosons: Many-body dynamics of bosonic systems, Phys. Rev. A {\bf{77}}, 033613 (2008).
\bibitem{stre} A.~I. Streltsov, O.~E. Alon, and L.~S. Cederbaum, Formation and Dynamics of Many-Boson Fragmented States in One-Dimensional Attractive Ultracold Gases, Phys. Rev. Lett. {\bf{100}}, 130401 (2008).

\bibitem{Reinh} D.~J. Masiello and W.~P. Reinhardt, Time-dependent quantum many-body theory of identical bosons in a double well: Early-time ballistic interferences of fragmented and number entangled states, Phys. Rev. A {\bf 76}, 043612 (2007).

\bibitem{pete} L. Cao, S. Krönke, O. Vendrell, P. Schmelcher, The multi-layer multi-configuration time-dependent Hartree method for bosons: Theory, implementation, and applications, J. Chem. Phys. {\bf 139}, 134103 (2013).

\bibitem{Scrinzi} J. Caillat, J. Zanghellini, M. Kitzler, O. Koch, W. Kreuzer, and A. Scrinzi, Correlated multielectron systems in strong laser fields: A multiconfiguration time-dependent Hartree-Fock approach, Phys. Rev. A {\bf 71}, 012712 (2005); J. Zanghellini, M. Kitzler, C. Fabian, T. Brabec, and A. Scrinzi, An MCTDHF approach to multielectron dynamics in laser fields, Laser Phys. {\bf 13}, 1064 (2003). 

\bibitem{Kato} T. Kato and H. Kono, Time-dependent multiconfiguration theory for electronic dynamics of molecules in an intense laser field, Chem. Phys. Lett. {\bf 392}, 533 (2004). 
\bibitem{Nest} M. Nest, T. Klamroth, and P. Saalfrank, The multiconfiguration time-dependent Hartree–Fock method for quantum chemical calculations, J. Chem. Phys. {\bf 122}, 124102 (2005). 
\bibitem{unified} O.~E. Alon, A.~I. Streltsov, and L.~S. Cederbaum, Unified view on multiconfigurational time propagation for systems consisting of identical particles, J. Chem. Phys. {\bf 127}, 154103 (2007).
\bibitem{haxton} D.~J. Haxton, K.~V. Lawler, and C.~W. McCurdy, Multiconfiguration time-dependent Hartree-Fock treatment of electronic and nuclear
dynamics in diatomic molecules, Phys. Rev. A {\bf 83}, 063416 (2011).
\bibitem{MCTDHF:He} D. Hochstuhl, S. Bauch, and M. Bonitz, Multiconfigurational time-dependent Hartree-Fock calculations for photoionization of one-dimensional Helium, J. Phys.: Conf. Ser. {\bf 220}, 012019 (2010).

\bibitem{exact} A.~U.~J. Lode, K. Sakmann, O.~E. Alon, L.~S. Cederbaum, and A.~I. Streltsov, Numerically exact quantum dynamics of bosons with time-dependent interactions of harmonic type, Phys. Rev. A {\bf 86}, 063606 (2012).  
\bibitem{exact2} A.~U.~J. Lode, {\it Tunneling Dynamics in Open Ultracold Bosonic Systems}, Springer Theses (Springer, Heidelberg, 2015). 
\bibitem{exact3} K. Sakmann, {\it Many-Body Schr\"odinger Dynamics of Bose-Einstein Condensates}, Springer Theses (Springer, Heidelberg, 2011).


\bibitem{MAP} A.~I. Streltsov, O.~E. Alon, and L.~S. Cederbaum, General mapping for bosonic and fermionic operators in Fock space, Phys. Rev. A {\bf 81}, 022124 (2010).

\bibitem{tunbos} A.~U.~J. Lode, A.~I. Streltsov, Kaspar Sakmann, O.~E. Alon, and L.~S. Cederbaum, How an interacting many-body system tunnels through a potential barrier to open space, Proc. Natl. Acad. Sci. USA {\bf 109}, 13521-13525 (2012).
\bibitem{ultr} A.~U.~J. Lode, M.~C. Tsatsos, and E. Fasshauer, {\it The Multiconfigurational Time-Dependent Hartree for Indistinguishable Particles X package (2015)}, http://mctdhx.org; http://ultracold.org; http://schroedinger.org; mctdh.bf.
\bibitem{IMP1} A.~I. Streltsov, Quantum systems of ultracold bosons with customized interparticle interactions, Phys. Rev. A {\bf 88}, 041602(R) (2013). 
\bibitem{IMP2}  O.~I. Streltsova, O.~E. Alon, L.~S. Cederbaum, and A.~I. Streltsov, Generic regimes of quantum many-body dynamics of trapped bosonic systems with strong repulsive interactions, Phys. Rev. A {\bf 89}, 061602(R) (2014).
\bibitem{IMP3} R. Beinke, S. Klaiman, L.~S. Cederbaum, A.~I. Streltsov, O.~E. Alon Many-body tunneling dynamics of Bose-Einstein condensates and vortex states in two spatial dimensions, arXiv:1508.03238 [cond-mat.quant-gas] (2015).

\bibitem{SupplMat} Supplementary Information available at \textbf{insert URL}
\bibitem{RJG} R.~J. Glauber, The Quantum Theory of Optical Coherence, Phys. Rev. {\bf 130}, 2529 (1963). 
\bibitem{RDMs} K. Sakmann, A.~I. Streltsov, O.~E. Alon, and L.~S. Cederbaum, Reduced density matrices and coherence of trapped interacting bosons, Phys. Rev. A {\bf 78}, 023615 (2008).


\bibitem{TDSE:1B} R. Kosloff, Time-dependent quantum-mechanical methods for molecular dynamics, J. Chem. Phys. {\bf 92}, 2087 (1988); W. H. Press, S. A. Teukolsky, W. T. Vetterling, and B. P. Flannery, \textit{Numerical Recipes in Fortran} (Cambridge University Press, Cambridge, England, 1992).

\bibitem{tunbos:old}  A.~U.~J. Lode, A.~I. Streltsov, O.~E. Alon, H.-D. Meyer, and L.~S. Cederbaum, Exact decay and tunnelling dynamics of interacting few-boson systems, J. Phys. B: At. Mol. Opt. Phys. {\bf 42}, 044018 (2009); ibid., Corrigendum, Phys. B: At. Mol. Opt. Phys. {\bf 43}, 029802 (2010).

\bibitem{tunbos2} A.~U.~J. Lode, S. Klaiman, O.~E. Alon, A.~I. Streltsov, and L.~S. Cederbaum, Controlling the velocities and the number of emitted particles in the tunneling to open space dynamics, Phys. Rev. A {\bf 89}, 053620 (2014).

\bibitem{socc1} I. B\v{r}ezinov\'{a}, A.~U.~J. Lode, A.~I. Streltsov, O.~E. Alon, L.~S. Cederbaum, and J. Burgd\"orfer, Wave chaos as signature for depletion of a Bose-Einstein condensate, Phys. Rev. A {\bf 86}, 013630 (2012).
\bibitem{barnali} A.~U.~J. Lode, B. Chakrabarti, V.~K.~B. Kota, Many-body entropies, correlations, and emergence of statistical relaxation in interaction quench dynamics of ultracold bosons, arXiv:1501.02611 [cond-mat.quant-gas] (2015).

\bibitem{socc}  M.~C. Tsatsos, and A.~U.~J. Lode, Vortex nucleation through fragmentation in a stirred resonant Bose-Einstein condensate, arXiv:1410.0414 [cond-mat.quant-gas] (2014). 

\bibitem{vort} S.~E. Weiner, M.~C. Tsatsos, L.~S. Cederbaum, A.~U.~J. Lode, Angular momentum in interacting many-body systems hides in phantom vortices, arXiv:1409.7670 [cond-mat.quant-gas] (2014).
\bibitem{prep} A.~U.~J. Lode and C. Bruder, in preparation.

\bibitem{Lewenstein94} M. Lewenstein, Ph. Balcou, M. Yu. Ivanov, A. L'Huillier, P.~B. Corkum, Theory of high-harmonic generation in low-frequency laser fields, Phys. Rev. A {\bf 49}, 2117 (1994).



\end{thebibliography}
\end{document}